\documentclass[twocolumn,showpacs]{revtex4}
\usepackage{graphicx}

\begin{document}

\preprint{APS/123-QED}

\title{Observation of Resonant Diffusive Radiation in Random Multilayered Systems}
\author{Zh.S.Gevorkian$^{1,2,*}$,S.R.Harutyunyan$^{3}$,N.S.Ananikian$^{1}$,V.H.Arakelian$^{2}$, R.B.Ayvazyan$^{1}$,\\
V.V.Gavalyan$^{1}$, N.K.Grigorian$^{1}$,H.S.Vardanyan$^{1}$,
V.H.Sahakian$^{1}$, A.A.Hakobyan$^{4}$}

\address{$^{1}$ Yerevan Physics Institute, Alikhanian Brothers St. 2, Yerevan 375036, Armenia.\\
$^{2}$ Institute of Radiophysics and Electronics, Ashtarak-2, 378410, Armenia.\\
$^{3}$ Institute for Physical Research
Ashtarak-2, 378410, Armenia. \\
$^{4}$  Byurakan  Astrophysical  Observatory, Byurakan, 378433,
Armenia.}

\begin{abstract}

Diffusive Radiation is a new type of radiation predicted to occur
in randomly inhomogeneous media due to the multiple scattering of
pseudophotons. This theoretical effect is now observed
experimentally. The radiation is generated by the passage of
electrons of energy $200KeV$-$2.2MeV$ through a random stack of
films in the visible light region. The radiation intensity
increases resonantly provided the $\check{C}$erenkov condition is
satisfied for the average dielectric constant of the medium. The
observed angular dependence and electron resonance energy are in
agreement with the theoretical predictions. These observations
open a road to application of diffusive radiation in particle
detection, astrophysics, soft X-ray generation and etc..

\end{abstract}

\pacs{41.60.-m, 07.85.Fv, 41.75.Fr, 87.59.-e}

  \maketitle

\indent{\it Introduction}. Several types of radiation mechanisms
are possible for a charged particle moving through a dielectric
medium. In the optical region well known mechanisms are the
isotropic luminescence, $\check{C}$erenkov radiation(CR),
bremsstrahlung and Transition radiation(TR). Bremsstrahlung is
related to the changing of particle velocity due to the collisions
with the atoms of the medium. In the optical region it is
negligible compared to other mechanisms. CR is produced when the
particle velocity exceeds the phase velocity of the radiation in
the medium and is emitted under a certain angle
$cos\theta=c/v\sqrt{\varepsilon}$, where $v$ is the velocity of
the particle and $\varepsilon$ is the dielectric constant of the
medium, see for example, \cite{Jelley}.

A charged particle uniformly moving in a randomly inhomogeneous
medium is known \cite{KaMi} to be radiating electromagnetic waves
due to the fluctuations of dielectric constant. The origin of the
radiation can be explained as follows. Each charged particle
creates an electromagnetic field around it which is not yet photon
but a pseudophoton. These pseudophotons  scatter on the
inhomogeneities of dielectric constant and convert into real
photons. The main issue here is accurately to take into account
the pseudophoton multiple scattering effects. The effect was
studied both for three dimensional (3D) \cite{ZH90} and one
dimensional randomness (1D) \cite{ZH98}. It was shown that the
radiation intensity consists of two terms. One is caused by the
single scattering of pseudophotons and the other by their multiple
scattering. When the conditions for multiple scattering are
fulfilled its contribution to the radiation intensity strongly
dominates. Single scattering contribution actually is the TR from
the randomly spaced interfaces.

 Random multilayered systems are an example of a medium with
1D randomness of dielectric constant. They are of significant
interest due to the possibility of employing them in high energy
particle detection \cite{Dolg}. Recently multilayered systems have
been considered as possible sources for soft X-ray radiation
\cite{Kap,Kn}. The CR and TR mechanisms were explored  for the
above mentioned purposes. Here we demonstrate the possibility of
experimental generation of a {\it new} diffusive radiation (DR)
mechanism in these systems. DR strictly differs from the other
radiation mechanisms by its angular distribution, dependence of
intensity on particle energy and etc.

For a given energy of electron the luminescence and CR photon
yields depend on the length of electron path in the medium and
does not related to the layered structure of the medium. Therefore
to reveal the stack effect we compare the photon yields from the
stack and a continuous medium with the same thickness of material,
see below.

{\it Theory}. Considering the radiation from a charged particle
uniformly moving in a system of randomly spaced parallel films it
was shown \cite{ZH98} that the multiple scattering of
pseudophotons leads to the diffusion and its contribution to the
spectral angular intensity is given by the formula
\begin{equation}
I(\omega,\theta)=\frac{5e^2\gamma_m^2(\omega) L_z l_{in}(\omega)
sin^2\theta} {2\varepsilon(\omega) c l^2 |cos\theta|} \label{abs}
\end{equation},

where $\varepsilon=\varepsilon_0+na(\varepsilon_f-\varepsilon_0)$
is the average dielectric constant of the system, $n$ is the
concentration of films, $a$ is their thickness, $\varepsilon_f$
and $\varepsilon_0$ are the dielectric constants of film and
medium, respectively,
$\gamma_m=(1-\frac{v^2\varepsilon}{c^2})^{-1/2}$ is the Lorentz
factor of the particle in the medium, $l$ and $l_{in}$ are elastic
and inelastic mean free paths of pseudophoton. The Eq.(\ref{abs})
is correct provided that $|cos\theta|\gg (\lambda/l)^{1/3}$ and
conditions for pseudophoton multiple scattering $\lambda\ll l\ll
l_{in}<L_z$, where $L_z$ is the system size in the $z$ direction,
are fulfilled. It is assumed that the $z$ axis is the normal to
the films and particle is moving on that direction. The formula
Eq.(\ref{abs}) has a clear physical meaning \cite{ZHNIM98}. The
quantity $e^2\gamma_m^2 L_z/c$ is the total number of
pseudophotons in the medium, $1/l$ is the probability of the
photon scattering and $l_{in}/l$ is the average number of the
pseudophoton scatterings in the medium. As it seen from
Eq.(\ref{abs}) the maximum of photon yield is achieved at the
resonant point $\gamma_m^{-2}=0$. Note that if one takes into
account the absorption of photons or finite sizes of the system
then the radiation intensity also will be finite although still
large. Close to the resonant point $\gamma_m^{2}$ is substituted
by the factor $d^2/\lambda^2$, where $d \equiv min[L,l_{in}]$,
$\lambda$ is the photon wavelength and $L$ is the characteristic
size of the system. The resonance condition $\gamma_m^{-2}=0$ is
the $\check{C}$erenkov condition for the average dielectric
constant. Hence one can say that the resonance diffusive radiation
(RDR) origins from the interaction of two processes. The
$\check{C}$erenkov  condition $v\sqrt{\varepsilon}/c=1$ creates
resonantly large number of pseudophotons and multiple scattering
converts them into real photons. It should be emphasized that in
the 3D random case the resonant factor $\gamma_m^2$ enters to the
radiation intensity as $ln\gamma_m$ \cite{ZH90}, therefore 1D
randomness is more preferable for getting larger photon yield.

It should be emphasized that the Eq.(\ref{abs}) is correct
provided that $v\sqrt{\varepsilon}/c\leq 1$. Note that the
resonant velocity $c/\sqrt{\varepsilon}$ is much larger than the
Cherenkov threshold velocity for the material
$c/\sqrt{\varepsilon_{f}}$, because
$\varepsilon<<\varepsilon_{f}$. Preliminary estimations show that
for electron energies far above the resonant energy DR is
suppressed as compared to CR.

Above we have taken into account absorption of pseudophotons in
the random medium. However already formed real photons also will
be absorbed in the medium. Therefore to know what part of already
created real photons will escape the system one should take into
account the absorption of real photons in the medium. Suppose that
we are interested in the photon yield from the depth $z$ in the
material. Using Eq.(\ref{abs}) and adding an exponential decaying
factor which takes into account the difference of paths of real
photons with different emitted angles, one has
\begin{equation}
\frac{dI}{dz}=\frac{5e^2\gamma_m^2(\omega)  l_{in}(\omega)
sin^2\theta} {2\varepsilon(\omega) c l^2
|cos\theta|}\exp\left(-\frac{z}{l_{in}|cos\theta|}\right)
\label{absrp}
\end{equation}
Here $\frac{dI}{dz}$ is the spectral-angular radiation intensity
per unit length of electron path in the medium. Note the
suppression of radiation intensity at very large angles. The real
DR photons are formed in the effective size $(ll_{in})^{1/2}$.
Therefore when finding the total radiation intensity one should
cut the integral  in the lower limit on this length. After
integration over the electron path, for the total spectral angular
intensity, we have
\begin{equation}
I=\frac{5e^2\gamma_m^2(\omega)  sin^2\theta l_{in}^2}
{2\varepsilon(\omega)
cl^2}\exp\left[-\left(\frac{l}{l_{in}}\right)^{1/2}\frac{1}{|cos\theta|}\right]
\label{absrpt}
\end{equation}

\begin{figure*}
 \includegraphics{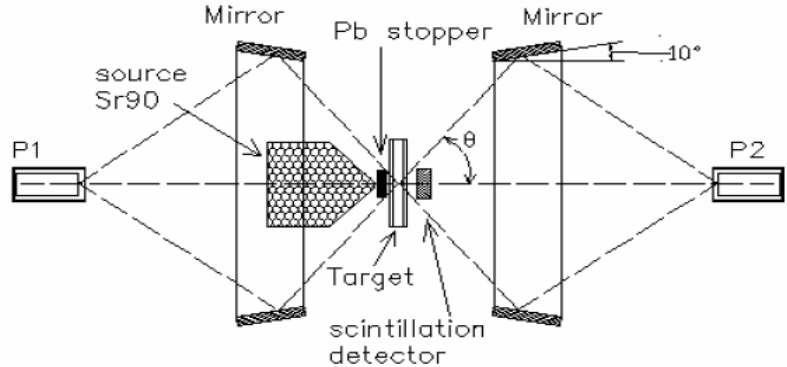}
  \caption{\label{fig1}Schematic of the experimental setup .}
\end{figure*}

So as one could expect the absorption of real DR photons leads to
the cutoff of radiation intensity at large angles and maximum lies
at medium angles.

{\it Experiment}.In the present work, we report the experimental
observation of the DR generated by electrons passing through a
random mylar stack. A schematic representation of the experimental
setup is shown in (Fig.~\ref{fig1}). The setup allows to detect
the photons in the wavelength range $300nm\div600nm$ and obtain
their angular distribution in the region $\theta=50^{\circ} \div
75^{\circ}$. The angular distribution is provided by the movement
of the photomultipliers $P1, P2$ along the horizontal axis of the
setup.

We use the following targets for generation of DR: A low-density
mylar stack consisting of $45$ films with thickness $4.5\mu m$ and
average spacing $55\mu m$; a dense mylar stack with $50$ films of
thickness $3\mu m$  and average spacing $6.8\mu m$ and a
polystyrene stack of $50$ films with thickness $3\mu m$ and
average distance between films $76\mu m$. The diameter of all
targets is $50mm$. For each stack it has been made a continuous
version with the same thickness of material as in the stack. As a
source of charged particles the radioactive $Sr^{90}$ giving
electrons of energy $0.2\div2.2MeV$ is used. In order to interrupt
the electron beam   a lead $Pb$ stopper placed between source  and
target is used. Electrons pass through the target reach the
scintillation detector and are absorbed there. The radiation
formed in the target is reflected and focused to the
photomultipliers $P1,P2$. The photomultipliers has sensitivity in
the wavelength region $300nm\div600nm$. The mirrors have a form of
truncated cone with reflecting aluminium coating in the inner
surface. The optical scheme of the setup allows to detect the
radiation in both forward and backward directions.The direction of
movement of electrons is assumed as a forward one. The photon
yields in both directions have been detected in the conventional
electron-photon coincidence technique. During  the experiment the
photomultipliers $P1,P2$ where moved step by step to the target
from the most distant position, which corresponds to minimal
detection angle $\theta=50^{\circ}$. The photon yields have been
registered after each  step of movement. The number of steps,
their widths and registration time were given by programmed
instruction. The number of steps ($3.4mm$ in widths) in the angle
region of $50^{\circ}\div75^{\circ}$ was $80$, that means at least
$3$ steps per degree.
 The data were analyzed by LabView program
package, collected by CAMAC and transferred to a PC by National
Instruments GPIB card.

Preliminary, the setup was calibrated under target which give an
isotropic radiation and acceptance curve of optical scheme was
obtained. As a source of an isotropic optical radiation we used a
scintillation polystyrene film with the same thickness as that of
targets. Then all obtained curves of radiation yield have been
normalized on the acceptance curve.

The intensity of the radiation yield in the optical range of
spectra can be represented as a sum of the following constituents:
luminescence, DR, $\check{C}$erenkov Radiation(CR) and Transition
Radiation. The luminescence contributes the essential part of the
radiation yield and spread both in the forward and backward
directions. The DR according to Eqs.(\ref{abs}-\ref{absrpt}) is
also the same for the backward and forward directions. The CR has
strictly forward direction for a continuous target and for stack,
because of the multiple scattering of real CR photons, spreads
both in forward and backward directions. The TR yield is
negligible as compared to the other constituents. The number of
optical TR photons per one electron is less than unity, $1/137$
per one interface, see for example \cite{GTs}, hence the TR
photons can not contribute to the electron-photon coincidence
events.

 The number of luminescence  photons for a given energy of
electron ought to be the same for the stack and the corresponding
continuous target with the same thickness of material. Therefore
to reveal the DR photons from the stack photon yield we extract
the photon yield of a continuous  medium with the same thickness
of the material.

\begin{figure}
 \includegraphics{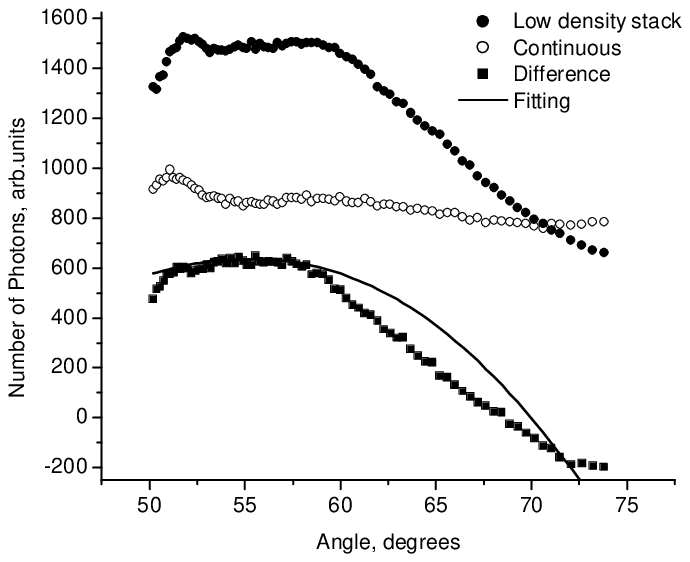}
  \caption{\label{fig2}Angular distribution of photons in the backward direction.}
\end{figure}

\begin{figure}
 \includegraphics{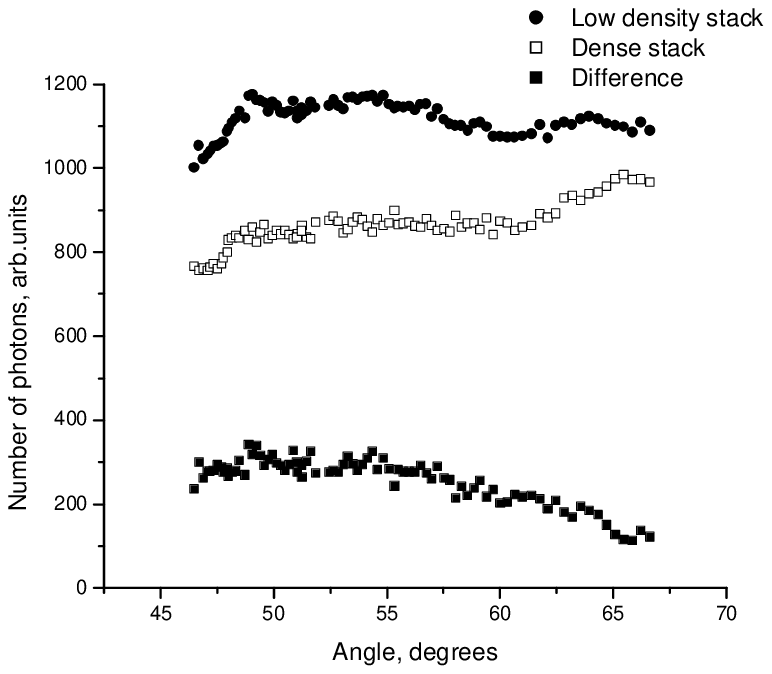}
  \caption{\label{fig3}Angular distribution of photons in the forward direction.}
\end{figure}

The obtained angular distribution of the low-density mylar stack
and the corresponding continuous target  are shown in
(Fig.~\ref{fig2})-(Fig.~\ref{fig3}). The photon yields from the
stacks are considerably higher than that of continuous medium. The
difference of these two yields we attribute to the presence of DR
radiation.

 Although the number of DR photons is the same for forward and
backward directions, in the experiment, the DR effect in back is
seen more clearly (Fig.~\ref{fig2}). The reason is the non proper
extraction of CR photons from the total photon yield. Apparently
the number of CR photons in the forward direction for the stack is
smaller than that for continuous target due to removing of some
photons from forward to back. Therefore the difference curve in
(Fig.~\ref{fig3}) provides underestimated values for DR photons.
In other hand the difference curve for backward direction
(Fig.~\ref{fig2}) provides overestimated values for DR photons
because of existence CR photons in backward for the stack. The
ratio of DR and CR intensities is of order $\lambda
\gamma_m^2/l$\cite{ZHNIM98}. This ratio is large near the electron
resonance energy. Therefore the DR photons give the dominant
contribution to the difference curve.

Suppose that there is no DR mechanism. In this case the photon
yield in the forward direction from the continuous target should
be larger than the yield from the stack because lost of some
forward photons due to reflections on the interfaces. Remind that
the number of possible TR photons in the stack is negligible.

The observed angular dependence of emitted DR photons in backward
direction (Fig.~\ref{fig2}) is in good agreement with the
theoretical relation Eq.(\ref{absrpt}). The maximum of DR
intensity is achieved approximately at $55^{\circ}$. One obtains
the same value from Eq.(\ref{absrpt}) provided that $l/l_{in} \sim
0.2$. If the parameter $l/l_{in}$ is smaller then the maximum
moves to the region of larger angles. We observed this phenomenon
checking as a target the polystyrene stack for which the maximum
is located at $65^{\circ}$. In the polystyrene the photon
absorption is weaker and correspondingly $l_{in}$ is larger.
Photon elastic mean free path $l$ in the geometrical optics region
$\lambda\ll a$ we are interested in is of order of average
distance between films \cite{ZH98}.

\begin{figure}
 \includegraphics{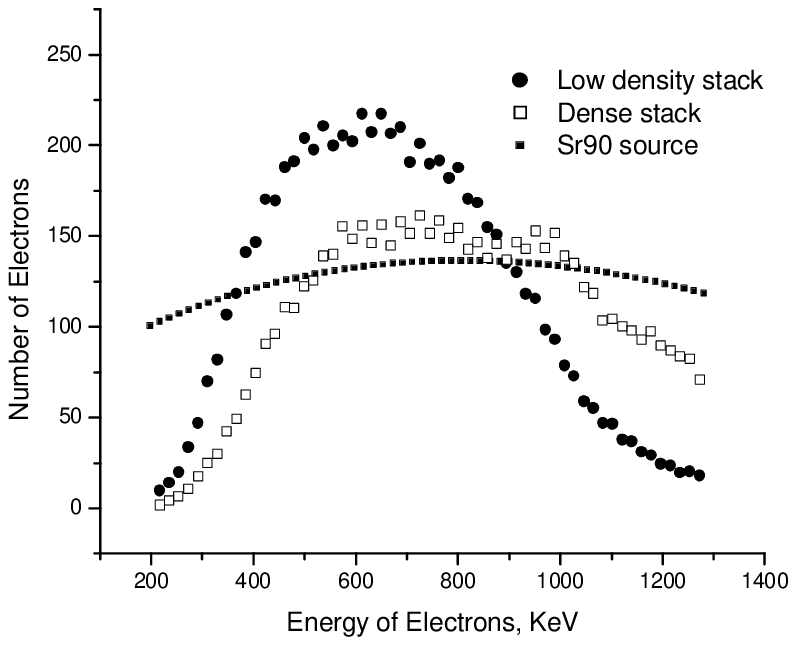}
  \caption{\label{fig4}Energy distribution functions of  radiating electrons
in electron-photon coincidence scheme.}
\end{figure}

We attempted to measure the energy of those electrons that radiate
DR photons. The adjacent averaged energy distributions for
different mylar targets (low-density stack, dense stack) and
electron source are presented in (Fig.~\ref{fig4}). The source and
low-density mylar stack distribution functions are normalized to
the same total number of electrons. They have maximum at the same
energy interval $0.6MeV\div1MeV$. However the heights of the
maximums are different. The highest maximum has the low-density
stack, second maximum has the dense stack and both exceed the
continuous mylar maximum. The coincidence of the place of peaks
from different targets is explained by the fact that the energy
distribution function of the electron source has a broad peak in
the region $600\div1000KeV$ \cite{source}, (Fig.~\ref{fig4}) . The
DR effect intensifies the height of stack maximums compared to the
continuous mylar maximum. For the low-density mylar stack with the
dielectric constant $\varepsilon_f=3.12$ of films, the average
dielectric constant approximately equals $1.17$ and the
corresponding resonance energy following from the condition
$v^2/c^2=1/\varepsilon$ is of order $750KeV$. This value lies in
the above mentioned interval of source maximum therefore a
resonant increasing of the height of the maximum takes place. The
resonance energy for the dense stack is of order $375KeV$
  and is far from the electron source maximum. Therefore the
increasing of height of the source maximum due to the DR effect is
less. Note that if the resonant factor $\gamma_m^2$ would not be
in the DR intensity Eq.(\ref{absrpt}) then the yield of photons
from the dense stack would be of the same order as from the
low-density stack because the ratio $l_{in}/l$ is the same for the
two stacks.

{\it Discussion}. Above  to reveal  the DR photons  we extract the
photon yield of a continuous target from the stack yield, assuming
that the luminescence from stack and continuous target is the
same. However the number of escaped luminescence photons  from
stack is smaller than that from continuous target because the
stronger absorption in the stack. This is caused by the scattering
on the interfaces and therefore longer photon paths in the medium.
The negative values in  the difference curve (Fig.~\ref{fig2}) is
also explained by the stronger absorption in the stack. We fit the
experimental data by the Eq.(~\ref{absrpt}) taking into account
the different absorption in the stack and continuous target, more
details see in \cite{detail}. As  follows from (Fig.~\ref{fig2})
the agreement in the maximum region is quite good. The discrepancy
at large angles between fitted and difference curves perhaps is
explained by the non proper normalization on the acceptance curve
for the stack.

We have observed Resonant Diffusive Radiation in the optical
region for the first time. The resonant character of radiation is
demonstrated. Angular dependence of the observed radiation and the
electron resonance energy value are in agreement with the theory.
The experimental confirmation of DR in the X-ray region could lead
to interesting applications in high energy particle detection
\cite{Zhchu}, in the soft X-ray generation and \cite{ZhVer} and
etc.

We are indebted to J.Verhoeven, M.J.van der Wiel , John Domingo,
A.Potylitsyn, V.Poghosov and A.Allahverdyan for helpful
discussions. This work was supported by a grant A-655 from
International Science and Technology Center. \footnote{*Electronic
address:gevork@yerphi.am}

\end{document}